\documentclass[a4paper]{article}

\usepackage{icrc2013}

\title{Highlights from the VERITAS Blazar Program}

\shorttitle{VERITAS Blazar Highlights}

\authors{
Jon Dumm$^{1}$ for the VERITAS Collaboration
}

\afiliations{
$^1$ University of Minnesota \\
}

\email{dumm@physics.umn.edu}

\abstract{
VERITAS is an array of four 12-m imaging Cherenkov telescopes, sensitive to gamma rays in the energy range from 85~GeV to 30~TeV.  VERITAS dedicates roughly 40\% of its total observing time to blazars. We present recent highlights from the VERITAS blazar discovery and long-term monitoring programs. These observations have led to 25 blazar detections, including 10 VHE discoveries.  The VERITAS cameras were recently upgraded, lowering the energy threshold and enhancing the sensitivity.  This upgrade helped to establish four new blazar detections with data taken after the completion of the upgrade in summer 2012.
}

\begin{document}
\maketitle

%Begin a section.
\section{Introduction}

Forty-five blazars have been detected at very high energy (VHE; $E>100$~GeV).  Blazars are a subclass of active galactic nuclei (AGNs) where the relativistic jet is oriented within a few degrees of the observer's line of sight \cite{Urry:1995mg}.  The broadband emission is generally accepted to be from a population of relativistic electrons accelerated somewhere in the jet, likely quite close to supermassive black hole where accretion is powering the jet.  The electrons synchrotron radiate from radio to X-ray and inverse-Compton is likely responsible for the gamma rays.  The emission is highly boosted towards the observer.  Although blazars are the least numerous type of AGN, blazars are the largest population of VHE-detected sources because of the enhanced detectability due to the boosting.  

VHE blazars are further broken down into four subclasses with detections of 34 high-frequency peaked BL Lacs (HBLs), 4 intermediate-frequency peaked BL Lacs (IBLs), and 4 low-frequency peaked BL Lacs (LBLs), as well as 3 flat spectrum radio galaxies (FSRQs).  The BL Lac classifications are largely based on the location of the synchrotron emission peak.  

The fluxes observed from these blazars generally range from 0.003 to 0.20 of the Crab Nebula flux (Crab Units, hereafter C.U.), but the fluxes can be quite variable, changing on timescales ranging from several minutes to years.  Just four blazars have been observed to produce over 1~C.U. with enhanced fluxes as short as several minutes or up to a few days.  Most spectra can be described by a power law over the VHE energy range with spectral indices ranging from about 2.5 -- 4.6.  Evidence for curvature in the spectrum is often seen in cases of strong detections.  The redshifts range from $z = 0.030$ to at least $z > 0.6035$, with PKS 1424+240 holding the record for the most distant spectrally-confirmed redshift (a lower limit) \cite{Furniss:2013roa}.  For such distant sources, the redshift- and energy-dependent imprint of VHE gamma-ray absorption on extragalactic background light (EBL) must be considered.  Recently, this effect has been quantified using H.E.S.S. blazar spectra \cite{Abramowski:2012ry}.  

Blazar observations give an opportunity for a large range of scientific goals.  We hope to understand properties of supermassive black holes and their environments, such as the requirements for launching relativistic jets and how they evolve with time.  Topics in fundamental physics include searches for new particles and processes such as Lorentz-invariance violation.  Finally, gamma-ray emission from blazars can increase our understanding of cosmology through indirect measurements of the EBL and intergalactic magnetic fields.  

\section{The VERITAS Blazar Program}

VERITAS has been collecting data with four telescopes since September 2007.  The array has an angular resolution (68\% containment) of $\sim$0.1$^{\circ}$, energy resolution of 15--25\%, and is able to detect a source with 1\% C.U. in $\sim$25 hours at 5 standard deviations ($\sigma$) \cite{Holder:2011fg}.  VERITAS is able to collect $\sim$1000 hours of data per year, $\sim$400 hours of which are dedicated to blazars.  All VHE blazars detected by VERITAS are shown in table~\ref{tab:blazars}.

\begin{table}[t]
\caption{\label{tab:blazars}
The 25 blazars detected at VHE with VERITAS. The 10 VHE discoveries are marked with $\dagger$. The reported redshifts are uncertain for B2 1215+30 ($z=0.130$ or 0.237) and 1ES 0502+675 ($z=0.341$). See text for redshift discussion of PG 1553+113 and 1ES 0647+250. The classifications based on synchrotron peak frequencies are taken from \cite{Nieppola:2005mz}, except for four cases: two (marked with asterisks) where the classification is determined from VERITAS-led MWL studies, and two (marked with $\beta$) where the historical HBL classification is used. 
}
\label{tab:sources}
\vspace{0.4cm}
\begin{center}
\begin{tabular}{c|c|c}
\hline
Source & Type & redshift (z) \\
\hline
Mrk 421 & HBL & 0.031 \\
Mrk 501 & HBL & 0.034 \\
1ES 2344+514 & HBL$^\beta$ & 0.044 \\
1ES 1959+650 & HBL & 0.047 \\
1ES 1727+502 & HBL & 0.055 \\
BL Lac & LBL & 0.069 \\
1ES 1741+196 & HBL & 0.083 \\
W Comae$^{\dagger}$ & IBL & 0.102 \\
VER J0521+211$^{\dagger}$ & IBL/HBL* & 0.108 \\
RGB J0710+591$^{\dagger}$ & HBL & 0.125 \\
H 1426+428 & HBL & 0.129 \\
B2 1215+30 & IBL/HBL & 0.130? \\
1ES 0806+524$^{\dagger}$ & HBL & 0.138 \\
1ES 0229+200 & HBL & 0.139 \\
1ES 1440+122$^{\dagger}$ & IBL/HBL & 0.162 \\
RX J0648.7+1516 & HBL* & 0.179 \\
1ES 1218+304 & HBL & 0.182 \\
RBS 0413$^{\dagger}$ & HBL & 0.190 \\
1ES 1011+496 & HBL & 0.212 \\
1ES 0414+009 & HBL & 0.287 \\
3C 66A$^{\dagger}$ & IBL & 0.3347 -- 0.41 \\
1ES 0502+675$^{\dagger}$ & HBL & 0.341? \\
PG 1553+113 & HBL$^\beta$ & 0.4 -- 0.62 \\
1ES 0647+250 & HBL & $\sim0.45$ \\
PKS 1424+240$^{\dagger}$ & IBL/HBL & $>$0.6035 \\
\hline
\end{tabular}
\end{center}
\end{table}

The VERITAS blazar key science plan was developed in 2010.  The plan splits the blazar observing time between obtaining deep exposures of known sources and discovery of new sources.  A certain amount of time is also dedicated to target-of-opportunity (ToO) programs.  In the latest observing season, the amount of time dedicated to known sources has increased to about 70\% of the total ($\sim$280 hours per year), and discovery efforts have been largely guided by ToO triggers.  
The scientific goals of the program are as follows:

{\bf i. Long-term Monitoring:}
VERITAS regularly monitors 14 selected VHE blazars, known as long-term plan (LTP) blazars.  This is about two-thirds of the known northern VHE BL Lac population.  Monitoring happens at regular intervals (generally once every three nights) while the source is highest to maximize the chances of detecting a flaring period.  We target anywhere from 5 to 25 hours per year on individual sources with the goal to build up deep archival exposures ($\sim$100--200 hours total).  

The targets were selected to represent a range of interesting topics.  Five of the LTP blazars are relatively distant, hard-spectrum sources, relevant for EBL/IGMF studies.  Four LTP blazars are nearby, bright HBLs which have exhibited extreme flares.  These flares are important for constraining emission mechanisms by studying flare evolution, distinguishing variable from constant cutoffs due to the EBL, and for searches for exotic phenomena like Lorentz-invariance violation.  
The last five LTP blazars are LBLs or IBLs where we aim to study the blazar sequence.  

Multiwavelength data are a key component to understanding blazars.  A large effort is made to obtain contemporaneous (and often truly simultaneous) radio, optical/UV, X-ray, and GeV data to allow for blazar modeling of the SEDs.  A number of ToO programs allow for us to quickly alert other facilities in the cases where flaring is detected.  

{\bf ii. Discovery Program:}
Originally motivated by radio--X-ray properties \cite{Costamante:2001ya}, the discovery program targets are now largely selected based on {\it Fermi}-LAT results.  While sources with marginal excess in previous VERITAS observations continue to be monitored, the focus has shifted to a smaller number of targets with high potential scientific payoff but often a smaller chance of detection.  For example, higher redshift candidates (relevant for EBL/IGMF) and non-HBLs to expand our knowledge to the rarer classes of BL Lacs.  

{\bf iii. ToO Program:}
ToO programs can be either self-triggered or externally triggered by a wide variety of instruments at any waveband.  Neutrino alerts from the IceCube detector can also trigger VERITAS observations.  Correlations between VHE and other wavelengths are routinely observed, so the purpose is to use the high duty cycle from the astronomical community to enable rapid response to unique opportunities with high scientific payout.  Partnerships with optical, X-ray, and other VHE observatories have now been in effect for several years.  

The sensitivity of VERITAS has been increased by several upgrades over the years.  In 2009, one of the telescopes was relocated, which increased the sensitivity by about 30\%.  An upgrade to the camera-level trigger was performed in 2011, allowing for triggering on shorter timescales.  Most recently, the old Photonis XP2970 photomultiplier tubes (PMTs) in the VERITAS cameras were replaced with super-bialkali PMTs (Hamamatsu R10560-100-20).  The new PMTs have a faster pulse profile and give $\sim$35\% higher yield for a Cherenkov spectrum resulting in a lower energy threshold for the array \cite{Kieda:2013}.  

\section{VERITAS Blazar Highlights}

% Recent results that may be worth adding: 
%RX J0648 & RBS 0413 -- noting that what might be interesting is the difficulty with SSC
%{\bf 1ES 0502+675} % Interesting enough?
% 1ES 0414+-009 -- again issues with SSC
%Fermi-6 (or Matteo's limits)

% 14-year study of Mrk 421
{\bf Mrk 421} was the first VHE-detected extragalactic source.  It is nearby at $z=0.031$ and generally is the brightest blazar in the VHE band.  Multiple outbursts have been observed to reach $\sim$10 C.U. allowing for very high signal-to-noise studies.  

Mrk 421 has been the subject of intense multiwavelength campaigns, including one this season between VERITAS and NuSTAR, as well as other telescopes.  During this latest campaign, the flux was generally less than 1~C.U. for several months before exhibiting an immense $\sim$10~C.U. flare, maintaining its bright state  above 1~C.U. for six nights.  Strong intra-night variability is observed (factor of $\sim$2 changes in flux), as well as strong night-to-night variability.  About 45 hours of VERITAS observations have been performed so far this season, and a large portion of that is truly simultaneous multiwavelength data, particularly with NuSTAR and Swift.  Analysis of this latest flaring event is underway.  A similar campaign for Mrk 501 is just starting.  

\begin{figure}[t]
  \centering
  \includegraphics[width=0.4\textwidth]{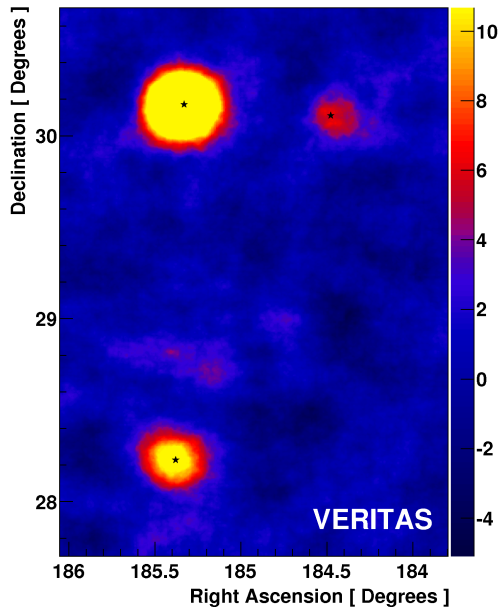}
  \caption{The significance map for a single field of view containing three VHE blazars (1ES 1218+304, W Comae and B2 1215+30). All three blazars are point-like but saturation on the color scale gives the appearance of different sizes. The total exposure for this map is $\sim$130 hours of observations, but the effective exposure varies considerably from one source to the next because the three sources were initially observed individually.}
  \label{fig:3sources}
 \end{figure}

% The 3-source map
{\bf B2~1215+30} is an IBL with uncertain redshift.  VERITAS observations between 2008 and 2012 detect the source at 8.9 $\sigma$ after 93 hours.  A significance map is given in Figure~\ref{fig:3sources}, which also shows nearby 1ES 1218+304 and W Comae contained in the same VERITAS field of view (3.5$^{\circ}$).  The flux is observed to change from between 1.2\% C.U. up to 3.4\% C.U. above 200~GeV on month-long timescales.  
 
On faster timescales, the source was observed to have both a low- and high-state in Swift UV and X-ray, leading to very different interpretations of the SED during each state.  VHE variability was not detected between these states.  Both states were successfully modeled by a synchrotron self-Compton model with the variability accounted for by changing the injection spectrum of electrons and the magnetic field strength.  Such a change could be explained by a change in the angle between the electrons and the magnetic field.  A detailed publication is underway.  

{\bf BL Lacertae}, the eponym of the BL Lac subclass of blazars, is at a redshift of $z=0.069$.  
It is an LBL that has been detected only sporadically at VHE gamma rays during flaring events.  VERITAS began more frequent monitoring of this source in June 2011 after observations of an enhanced state from {\it Fermi}-LAT and AGILE.  

On 2011 June 28, VERITAS detected a bright flaring event from BL Lacertae with $\sim$125\% C.U. peak flux.  The spectrum is well described by a power law with index $\Gamma = 3.6 \pm 0.4$, which is consistent with an earlier flare reported by MAGIC.  Even though the rising edge of the flare was not observed, the short exposure (34.6 min.) allowed measurement of an exponential decay constant $\tau = 13 \pm 4$ min.  This is the first report of such rapid variability in an LBL.  The rapid timescale constrains the emission region to be quite small.  The VHE flare is tied to the emergence of a new radio knot resolved to be moving down the jet.  A strong radio flare four months later could also be linked to the VHE flare \cite{Arlen:2012mm}.

% Flare with coincident Swift data
{\bf VER J0521+211} was observed for 4 hours in 2012 Nov 11--14. The source was clearly detected at a flux level of $\sim$10\%~C.U., in between that of flaring and non-flaring states observed in 2009.  
%about a factor of two higher than that of the discovery observations of the source in 2009. 
Simultaneous X-ray observations with Swift-XRT were obtained, and a detailed analysis of the multiwavelength SED is in preparation.

{\bf 3C66A and W Comae} are both IBLs, a rare class of VHE blazars.  So far, the low- and intermediate-frequency-peaked BL Lacs (LBLs and IBLs) are mainly detected during flaring episodes.  Overall, LBLs are more powerful and more luminous than HBLs.  Jet alignment is one possible explanation for shifts in the observed peak frequency.  Other scenarios involve accretion rate differences \cite{Meyer:2011uk}.  

Since blazar spectra have been observed to harden during flaring states, low-state detections of LBLs and IBLs are needed to construct unbiased SEDs.  Only with low-state detections can we accurately explore relations between the low- and high-frequency peaks in blazar SEDs.  Recently, VERITAS was able to accumulate enough observation time to detect two IBLs, W Comae and 3C66A, in states where the flux is about five times lower than previous detections during flaring episodes.  Detailed multiwavelength analysis on these data are underway.  

{\bf 1ES 1440+122} is a BL Lac at $z=0.162$.  The object was discovered by VERITAS at a significance level of 5.5 $\sigma$ during the 2008--2010 observing seasons.   Combining data from Swift X-rays, {\it Fermi}-LAT high energy gamma rays, and VERITAS VHE gamma rays, SEDs were fitted finding optimal jet parameters for synchrotron self-Compton, external-Compton, and hadronic models.  All models reproduce the data well, resulting in jet parameters in line with other BL Lacs.  An external radiation field is necessary to get  close to equipartition between the relativistic electron population and the magnetic field.  

1ES 1440+122 was originally classified as an IBL \cite{Nieppola:2005mz} or a borderline HBL.  The VERITAS observations and multiwavelength modeling efforts argue for firm classification as a HBL, with the synchrotron emission peaking around $10^{17}$~Hz.  

{\bf PG 1553+113} is an HBL with uncertain redshift.  The redshift does have a firm lower limit of $z>0.40$ \cite{Danforth:2010vy}, making it one of the most distant VHE-detected blazars.  It has a hard spectrum as seen by the {\it Fermi}-LAT ($\Gamma_{\mathrm{LAT}} = 1.67 \pm 0.022$) \cite{Fermi-LAT:2011iqa}.  This hard spectrum for lower energy gamma rays stands in contrast to the very soft spectrum observed at VHE ($\Gamma_{\mathrm{VHE}} = 4.3 \pm 0.1$) where attenuation on extragalactic background light (EBL) is evident.

VERITAS performed deep observations of PG 1553+113, accumulating 80 hours of quality-selected data between 2010 and 2012 and resulting in the highest VHE signal-to-noise spectrum yet.  The time-averaged flux during this period corresponds to 6.9\%~C.U. above 160~GeV.  There is evidence for long-term variability in the flux, which reached as high as 18.3\%~C.U. above 160~GeV.

The gamma-ray spectrum of PG~1553+113 can be used to constrain the source distance by excluding redshifts that would produce an EBL-corrected spectrum with an upturn towards higher energies, a feature not observed in any other VHE blazar \cite{Mazin:2006ww}. Using a minimal EBL model to derive the intrinsic gamma-ray spectrum, an upper limit of $z<0.62$ is found from the VERITAS data.

{\bf 1ES 0229+200} is a relatively distant ($z=0.1396$) HBL with a hard spectrum ($\Gamma_{VHE} \sim 2.5$) and a flux of $\sim$2\%~C.U. above 300~GeV.  Over a period of three years, VERITAS collected 54.3 hours on this source, resulting in a strong 11.7 $\sigma$ detection.  These data provide the first evidence of variability at VHE for this blazar on year-long timescales (1.6\% chance of being constant over the 3-year period).  This variability makes use of 1ES 0229+200 for studies of the intergalactic magnetic field (IGMF) more difficult since they now must cope with inherent uncertainties from time-averaging SEDs.  It also might challenge models where the VHE photons are secondaries produced by the interaction of primary cosmic rays ($E=10^{16}$--$10^{19}$~eV) with background photons \cite{Essey:2010er} because the reprocessed emission is not expected to show such short-scale time variability \cite{Prosekin:2012ne}.  

{\bf PKS 1424+240} is an IBL with a lower limit redshift of $z>0.6035$ \cite{Furniss:2013roa}.  This recently published limit makes it the most distant spectrally-confirmed VHE source.  VERITAS spectral measurements extending up to 500~GeV probe the EBL for large optical depths ($\tau$), at least $\tau>4$, for minimal EBL models, and up to $\tau=5$.  The new redshift results prompted VERITAS to take a deeper exposure on this source, accumulating about 76 quality-selected hours so far this season with analysis underway.  

%After correcting for EBL absorption, the VERITAS spectrum is lower than expected from a smooth extrapolation of the {\it Fermi}-LAT spectrum.  Only at 500~GeV does the VERITAS spectrum begin to match the extrapolation from {\it Fermi}-LAT results \cite{Furniss:2013roa}.  

{\bf 1ES~1741+196} was originally discovered at VHE by MAGIC in 2011 and reported to have a very small flux of 0.8\%~C.U. \cite{Berger:2011sy}.  This HBL at redshift z=0.083 was detected by VERITAS for the first time at 5.3 $\sigma$ after 26.8 quality-selected hours of observation, resulting in an integral flux of (1.3$\pm$0.4)\% C.U.  Multiwavelength analysis is currently underway.  

{\bf 1ES~0647+250} was previously detected by MAGIC in 2011 and was reported to have an integrated flux of 3\% C.U. \cite{1742-6596-375-5-052021}.  The redshift is not well measured spectroscopically, but three imaging redshift estimates are available: 
$z = 0.45 \pm 0.08$ \cite{Meisner:2010wi}, $z = 0.41 \pm 0.06$ \cite{Kotilainen:2011tw}, and  $z > 0.49$ \cite{Shaw:2013pp}.  

The source was selected as a VHE candidate on the basis of its radio, optical, and X-ray properties \cite{Costamante:2001ya} but also because of the hard high-energy gamma-ray spectrum from the {\it Fermi}-LAT ($\Gamma_{\mathrm{LAT}} = 1.59 \pm 0.08$).  VERITAS began observing the source during partial moonlight with raised trigger thresholds.  After noting a possible enhanced state compared to previous observations, a target-of-opportunity program was initiated to get 10.9~hours of dark time, yielding a detection at 6.2 $\sigma$ with a flux of $(2.9\pm0.7)$\%~C.U. above 140~GeV, in close agreement with the MAGIC detection.  It has a soft spectrum with $\Gamma_{\mathrm{VHE}} = 3.6 \pm 0.8$.  This is the first VERITAS detection of the source.  Multiwavelength data were collected and analysis is ongoing.  

{\bf 1ES~1011+496} was observed to be in a higher-than-normal state the same night as 1ES 0647+250.  Target-of-opportunity observations collected 10.4 hours of quality-selected data, yielding an 8.5 $\sigma$ detection.  This is also the first time VERITAS detected the source.  The flux was observed to be $(6.3\pm1.4)$\%~C.U. above 150~GeV, in rough agreement with the MAGIC-reported flux of 7\%~C.U. above 200~GeV \cite{Albert:2007kw}.  The spectral index was observed to be $\Gamma_{\mathrm{VHE}} = 3.0 \pm 0.3$.  Multiwavelength data were collected and analysis is ongoing.  

% New discovery in Nextday
{\bf 1ES~1727+502} is an HBL at redshift $z=0.055$.  It was first detected at VHE by MAGIC and reported to have a flux of $(2.1 \pm 0.4)$\%~C.U. above 150~GeV with no indications of variability.  The spectral index for a power-law fit is $\Gamma_{\mathrm{VHE}} = 2.7 \pm 0.5$ \cite{Aleksic:2013fga}.  VERITAS observed the source during partial moonlight, resulting in a strong detection, which indicates an enhanced flux over the MAGIC-reported values.  This is the first VERITAS detection of the source and further analysis is underway.  

%Upper limits 
VERITAS has observed more than 100 blazars from 2007 to 2012, and the majority of them have not been detected ($>5 \sigma$) at VHE. A study of this data set, including all the 2FGL-catalog sources in the field of view of VERITAS observations, is currently ongoing in order to estimate the upper limits on the VHE emission of each source. We also compute the results from a stacked analysis of the observations, for the entire data set and for smaller subsets, defined as a function of the source redshift and blazar class. Preliminary results show a stacked significance of ~$4 \sigma$, most likely associated with nearby HBLs \cite{Cerruti:2013}.  

\section{Conclusions}
Blazar observations are a large component ($\sim$40\%) of the VERITAS observing program.  The long-term plan includes both producing unprecedented VHE data sets on known sources as well as discovery observations.  In the last season, this effort has shifted in favor of building up deep observations of known sources ($\sim$70\% of the time).  Even so, with the upgrade of the PMTs in the camera and monitoring of candidate sources during partial moonlight time, we were able to confirm four previous MAGIC detections for the first time.  In conjunction with VHE data collection, we have an organized effort in place to collect multiwavelength data.  These data allow us to address the big-picture questions in supermassive black holes, fundamental physics, and cosmology.  

\vspace*{0.5cm}
\footnotesize{{\bf Acknowledgment:}{This research is supported by grants from the U.S. Department of
Energy Office of Science, the U.S. National Science Foundation and the Smithsonian
Institution, by NSERC in Canada, by Science Foundation Ireland
(SFI10/RFP/AST2748) and by STFC in the U.K. We acknowledge the excellent
work of the technical support staff at the Fred Lawrence Whipple Observatory and
at the collaborating institutions in the construction and operation
of the instrument.}}

\bibliographystyle{unsrt}    % for BibTeX - sorted numerical labels by order of first citation.
\bibliography{myrefs}           % expects file "myrefs.bib"

\end{document}